%% file: main.tex
\documentclass[10pt]{article}

\usepackage[margin=0.78in]{geometry}
\usepackage{amsmath,amssymb}
\usepackage{booktabs}
\usepackage{graphicx}
\usepackage[hidelinks]{hyperref}
\usepackage{microtype}
\usepackage{xcolor}
\usepackage{tabularx}
\usepackage{url}
\usepackage{authblk}

\title{Evaluation-Recording Contamination in Learned
Nano-Quadrotor Dynamics:\\A Fresh-Seed Audit}
\author[]{David Shulman\thanks{Corresponding author. Email:
\href{mailto:david.shulman.research@gmail.com}{
\texttt{david.shulman.research@gmail.com}}}}
\affil[]{Department of Statistics, University of Haifa, Haifa, Israel}
\date{}

\input{generated_results}

\input{manuscript_extensions}

\begin{document}
\maketitle

\begin{abstract}
Learned flight-dynamics models are commonly trained on short windows cut from
longer recordings. Randomly assigning those windows to fitting and evaluation
can place strongly dependent samples from one physical flight on both sides
of the split. We audit this practice on a pinned snapshot of the NanoBench
Crazyflie 2.1 dataset. The controlled experiment holds evaluation flights,
rollout starts, fitting-window count, validation data, and optimization budget
fixed. A contaminated protocol replaces \LeakagePercent\% of a
recording-disjoint fitting set with windows from the evaluation recordings;
both arms use the same clean recording-disjoint validation set. We compare a
world-coordinate delta multilayer perceptron (MLP) with a
relative-coordinate control over \NumSeeds\ fresh, precommitted training seeds
and \NumEvalFlights\ complete evaluation recordings. The sole primary
endpoint is 1\,s failure-aware position root-mean-square error (RMSE), with
paired crossed recording--seed inference. Contamination lowers apparent
world-MLP error by 12.1\%, from 0.299 to 0.262\,m, but the predeclared 95\%
interval for contaminated minus disjoint error is
$[-0.0735,0.0011]$\,m. The effect-size threshold is met but the interval
crosses zero, so the confirmatory result is negative. Secondary horizons and
the relative-coordinate model show lower contaminated-arm error, yet a direct
coordinate-representation interaction is unsupported. The audit demonstrates
that leakage claims in flight modeling require recording-level splits,
failure-aware rollouts, fresh-seed replication, and inference over both
physical recordings and optimization seeds.
\end{abstract}

\section{Introduction}

Data-driven dynamics models are now routinely used for system identification,
prediction, adaptive control, and model-predictive control of aerial vehicles
\cite{bansal2016learning,torrente2021ddmpc,hewing2020lbmpc}. Their utility
depends not only on model capacity but on the meaning of ``unseen'' in the
evaluation. A flight log is not a collection of independent rows. Adjacent
samples share vehicle state, reference commands, control history,
disturbances, battery condition, motion-capture frame, and preprocessing.
Windows taken from the same recording can therefore remain strongly related
even when their exact sample indices differ.

This dependence creates a particular form of evaluation leakage. A random
window split can measure interpolation within familiar physical runs rather
than generalization to a new flight. The numerical test score may be valid for
the random-window estimand while being misleading for deployment on a future
recording. Similar discrepancies between sampled rows and intended deployment
units are documented in machine-learning science, hierarchical data, clinical
prediction, and time series
\cite{kapoor2023leakage,kaufman2012leakage,roberts2017crossvalidation,
saeb2017usecase,bergmeir2018timeseries}. The issue is especially consequential
in flight: a dynamics model can have low one-step loss yet drift or diverge
when recursively rolled forward.

NanoBench provides synchronized actuator commands, vehicle state, controller
internals, and motion-capture measurements for a 27\,g Crazyflie 2.1
\cite{ullah2026nanobench}. Its repeated trajectories make it useful for
testing evaluation design. This paper does not propose a flight controller or
a state-of-the-art dynamics architecture. It asks a narrower methodological
question that precedes controller design:

\begin{quote}
\emph{How much does fitting-time access to evaluation recordings change
open-loop dynamics scores when window count, clean validation data, rollout
starts, architecture, and optimization budget are controlled?}
\end{quote}

The study began with an exploratory result that appeared strongly positive.
Independent review identified training-set-size imbalance, contaminated
validation, post hoc seed expansion, coordinate-feature confounding, and
rollout censoring as threats to that conclusion. We therefore locked a new
protocol before running seeds 100--119 and treat the earlier experiment only
as hypothesis generation. The distinction is material: the corrected primary
experiment is negative under its preregistered decision rule.

The contributions are:
\begin{enumerate}
  \item a machine-audited comparison that differs only in declared
  evaluation-recording access during fitting;
  \item fixed complete-recording evaluation with explicit rollout starts;
  \item a failure-aware metric that retains divergent trials rather than
  silently conditioning on successful rollouts;
  \item crossed inference over physical recordings and fresh training seeds;
  \item a translation-relative model and a direct difference-in-differences
  test of coordinate moderation; and
  \item a reproducible record of the exploratory-to-confirmatory chronology,
  including the negative confirmatory decision.
\end{enumerate}

\section{Related work}

\subsection{Leakage, grouped validation, and the deployment unit}

Leakage broadly occurs when information unavailable under the intended use
case enters fitting, model selection, preprocessing, or evaluation
\cite{kaufman2012leakage,kapoor2023leakage}. It is not restricted to literal
duplicate rows. Dependency-aware splitting methods group structurally related
samples so that the held-out unit resembles the future prediction unit
\cite{roberts2017crossvalidation,joeres2025datasail}. The same principle
appears in clinical machine learning, where patient-level random splits can
answer a different question from prediction for a new patient
\cite{saeb2017usecase}.

Cross-validation is an estimator for a specified resampling regime, not a
universal property of a model. Its uncertainty can be large when the number
of independent units is small \cite{varoq2018cvfailure}, and repeated model
selection against the same validation data can induce a second layer of
optimism \cite{cawley2010overfitting}. Surveys of cross-validation emphasize
that exchangeability assumptions and the target estimand must be stated
\cite{arlot2010survey,bates2024crossvalidation}. For temporally ordered data,
random folds may sometimes estimate an interpolation objective, but blocked
or forward designs are required for many forecasting objectives
\cite{bergmeir2018timeseries,cerqueira2020forecast}. Consequently, this audit
does not claim that every random-window split is mathematically invalid. It
tests whether same-recording access changes scores for the explicit target of
generalizing to a new flight recording.

\subsection{Learned quadrotor dynamics}

Classical system identification separates model structure, excitation,
estimation, and validation \cite{ljung1999system}. Crazyflie-specific work has
shown that even nano-quadrotors admit useful identified models, while also
highlighting actuator, aerodynamic, and sensing constraints
\cite{foerster2015crazyflie}. Neural dynamics models can capture effects that
are cumbersome to specify analytically
\cite{punjani2015helicopter,bansal2016learning}. Hybrid aerodynamic learning
further combines physical structure and learned residuals
\cite{bauersfeld2021neurobem}, and learned models have supported aggressive
model-predictive flight \cite{torrente2021ddmpc} and rapid adaptation in wind
\cite{oconnell2022neuralfly}.

These advances make evaluation design more, not less, important. Recursive
prediction compounds one-step errors, while closed-loop control introduces
distribution shift and safety constraints. Reviews of learning-based control
therefore distinguish predictive accuracy from closed-loop guarantees
\cite{hewing2020lbmpc,brunke2022safelearning}. Our experiment occupies the
earlier system-identification stage: it evaluates open-loop state prediction
under measured motor inputs. The result should inform benchmark construction,
but it does not establish controller stability or safe hardware operation.

\subsection{Crossed sources of uncertainty and interaction tests}

Each error difference in this study is indexed by an evaluation recording and
a training seed. Treating 520 recording--seed cells as independent would
pseudoreplicate both sources. Crossed resampling and variance methods were
developed for array-structured data with multiple random factors
\cite{owen2007pigeonhole,owen2012arrays}. We use a transparent two-way
random-effects decomposition and conservatively anchor interval degrees of
freedom to the 20 fresh seeds.

The coordinate-control question also requires an interaction. The fact that
one model has an interval excluding zero while another does not is not itself
evidence that their effects differ \cite{gelman2006difference,
nieuwenhuis2011interaction}. We therefore estimate the direct difference
between the two paired contamination gaps.

\section{Dataset and prediction task}

\subsection{Pinned NanoBench snapshot}

NanoBench is a multi-task benchmark for nano-quadrotor identification,
control, and state estimation \cite{ullah2026nanobench}. We use the flight-log
component for one-step dynamics training and open-loop rollout evaluation,
which is closest to its system-identification objective. The neural
architecture is adapted from the upstream learned-dynamics implementation
used by the control benchmark; the present split audit and statistical
analysis are separate contributions.

The pinned data directory contains 107 CSV recordings sampled at 100\,Hz.
The declared inclusion rule requires at least 200 active samples with
motor-1 PWM greater than 1000. It retains 106 recordings and excludes
\texttt{B7\_oval\_slow\_rep1}, which contains 211 rows but no active row under
that rule. This accounting is intentionally snapshot-specific; a broader
dataset release claim must not be substituted for the exact files analyzed.
A row-level audit records every included and excluded file.

\subsection{State, action, and one-step target}

At sample $t$, the 14-dimensional state is
\begin{equation}
x_t =
\left[
p_t^\top,\;
v_t^\top,\;
q_t^\top,\;
r_t^\top,\;
b_t
\right]^\top ,
\end{equation}
where $p_t\in\mathbb{R}^3$ is Vicon world position,
$v_t\in\mathbb{R}^3$ is linear velocity, $q_t\in\mathbb{R}^4$ is a unit
quaternion, $r_t\in\mathbb{R}^3$ is reference position, and $b_t$ is battery
voltage. The action $u_t\in[0,1]^4$ contains normalized motor commands. The
network predicts a standardized state increment,
\begin{equation}
\widehat{\Delta z}_t=f_\theta(\widetilde{x}_t,u_t),\qquad
\hat{x}_{t+1}=S^{-1}\!\left(S(x_t)+\widehat{\Delta z}_t\right),
\end{equation}
where $S$ is fitted from the corresponding arm's fitting data only.
Increment targets for the exogenous reference and battery components are set
to zero. During rollout those four components are overwritten by their
measured next-sample values. Thus the experiment tests vehicle-state
prediction conditional on the recorded command, reference, and battery
sequence; it is not a reference-trajectory generator or motor controller.

\subsection{Evaluation recordings and condition composition}

A fixed split seed selects \NumEvalFlights\ complete evaluation recordings
from the 106 usable recordings. Fifty valid starts per recording are stored as
explicit (flight ID, sample index) pairs, giving 1,300 registered rollouts per
model, protocol, and training seed. Every start has at least 200 subsequent
samples, enabling horizons of 0.5, 1, and 2\,s. The identical starts are used
in every comparison.

The evaluation set is heterogeneous but not balanced by condition:
12 of 26 recordings are trefoils, four are helices, three are circles, two
are figure-eights, two are ovals, and one each is a star, random-waypoint, and
multisine system-identification flight. Metadata labels 21 recordings
``slow,'' four ``medium,'' and the multisine speed as unknown. This
composition motivates both flight-level inference and the prespecified
trajectory-type--speed grouping sensitivity. It also limits generalization:
the study estimates a split-seed-42 mixture, not an equal-weighted universe of
all maneuver families.

\section{Experimental design}

\subsection{Exploratory chronology and confirmatory lock}

The first-stage exploratory experiment used ten inspected seeds and produced a
large apparent gap. Independent audit then identified that its contaminated
and safe arms did not have comparable recording support and that validation
could share evaluation recordings. The relative-coordinate ablation also
removed both absolute position and the reference, eliminating tracking
information rather than only removing the arbitrary world origin.

Before any seed in the confirmatory set was executed, the repository locked:
the split seed, evaluation starts, fitting and validation counts,
contamination fraction, models, seeds 100--119, primary endpoint, crossed
interval, and two-part decision rule. A later performance-only amendment
cached immutable CSV arrays after timing one partial run; no scientific metric
had been inspected, and all seeds were restarted. The preregistration and
amendment remain versioned with the code. All results in this manuscript come
from the restarted confirmatory directory. Exploratory numbers are excluded
from inference.

\subsection{Recording-disjoint base and shared validation}

Let $E$ denote the 26 evaluation recordings and $R\setminus E$ the remaining
recordings. From $R\setminus E$, the procedure draws 20,000 clean one-step
windows. Eight complete non-evaluation recordings, chosen with validation
seed 143, provide a shared 2,000-window validation set. Another 2,000 windows
from those validation recordings are discarded so that validation remains
grouped while the fitting count stays fixed. The remaining 16,000 windows
form the recording-disjoint fitting set $T_{\mathrm{safe}}$.

The contaminated arm begins with that exact set. It removes 4,000 uniformly
selected clean windows and replaces them with 4,000 windows from $E$:
\begin{equation}
T_{\mathrm{contam}}
=
\left(T_{\mathrm{safe}}\setminus D\right)\cup C,\quad
|D|=|C|=4000.
\end{equation}
Therefore both arms contain \NumFitWindows\ fitting windows and share
\SharedBasePercent\% of their window identities. They use the identical
\NumValidationWindows-window clean validation manifest. No validation
recording appears in $E$, and no evaluation window determines early stopping.
This construction isolates fitting-time evaluation-recording access while
holding fit volume, validation evidence, optimization settings, and rollout
evaluation fixed.

\begin{table}[ht]
\centering
\caption{Controlled comparison. ``Same'' means identical stored window or
rollout manifests, not merely equal counts.}
\label{tab:design}
\small
\begin{tabularx}{\linewidth}{lXX}
\toprule
Component & Recording-disjoint arm & Contaminated arm \\
\midrule
Fitting windows & 16,000 clean windows & 12,000 shared clean + 4,000 from
evaluation recordings \\
Validation & \multicolumn{2}{c}{Same 2,000 windows from eight
non-evaluation recordings} \\
Evaluation & \multicolumn{2}{c}{Same 26 recordings and same 1,300 rollout
starts} \\
Normalization & Fit on arm's own fitting windows & Fit on arm's own fitting
windows \\
Optimization & \multicolumn{2}{c}{Same architecture, budget, seed, and early
stopping rule} \\
\bottomrule
\end{tabularx}
\end{table}

\subsection{Treatment-strength audit}

Exact registered start indices are excluded when drawing $C$. That exclusion
does not imply temporal independence: neighboring samples from the same
flight share nearly all relevant history. The post-run treatment audit finds
that 3,620 of the 4,000 contaminating windows (90.5\%) fall within at least
one evaluated 2\,s segment, and the median distance to the nearest registered
start is 26 samples, or 0.26\,s. The treatment is thus a deliberately strong
stress test---near-complete fitting access to evaluated trajectory
segments---rather than a faithful reimplementation of every conventional
random-window split. This wording is important for external validity.

\section{Models and optimization}

\subsection{World and relative inputs}

Both learned models use three fully connected hidden layers of 256 ReLU units
and a 14-dimensional linear output:
\begin{equation}
f_\theta:\mathbb{R}^{18}\rightarrow\mathbb{R}^{14},\qquad
h_{k+1}=\operatorname{ReLU}(W_kh_k+c_k).
\end{equation}
The world model receives the standardized $x_t$ concatenated with $u_t$.
The relative model replaces the network-input position with zero and the
reference with tracking error:
\begin{equation}
\widetilde{x}^{\mathrm{rel}}_t =
\left[
0_3^\top,\;v_t^\top,\;q_t^\top,\;(r_t-p_t)^\top,\;b_t
\right]^\top .
\end{equation}
Its physical target and recursive state remain in world coordinates. A joint
translation $p_t\mapsto p_t+c$, $r_t\mapsto r_t+c$ therefore leaves the
network features unchanged while the predicted world position translates
with its integration origin. This is a genuine relative-coordinate control,
unlike the exploratory ablation that discarded reference error.

\subsection{Training protocol}

Each arm fits its own state standardizer on fitting data only. The
relative-input model additionally fits an input standardizer to its
transformed features. The loss is mean squared error on standardized state
increments. Adam uses learning rate $10^{-3}$, batch size 1,024, and
Gaussian action noise with standard deviation 0.02 clipped to $[0,1]$.
A cosine schedule spans at most 100 epochs. Early stopping with patience 15
uses the shared clean grouped validation set. For a given model and seed, the
two arms use the same training seed, producing a paired optimization
comparison. No seed is removed based on its performance.

The architecture is intentionally compact and close to the upstream learned
dynamics network. Persistence and constant-velocity predictors exist in the
software as parameter-free checks: because they fit no split-dependent
parameters, their output cannot demonstrate or refute learned contamination
bias. We therefore do not promote their zero protocol gap as scientific
evidence. Physics-plus-residual and higher-capacity architectures would test
model-class generality but were not added after inspecting the locked result.

\section{Evaluation metrics and statistical analysis}

\subsection{Open-loop rollouts and failure-aware error}

At each registered start the model is rolled open loop using measured future
motor commands, references, and battery voltage. Position, velocity, attitude,
and divergence are evaluated at 0.5, 1, and 2\,s. Quaternion attitude error
uses the shortest rotation angle. A rollout is marked divergent when position
error first exceeds 2\,m or becomes non-finite.

Conventional RMSE calculated only on finite surviving rollouts can improve
when the most difficult trials vanish. The primary failure-aware position
metric retains all $N$ registered starts:
\begin{equation}
\operatorname{RMSE}_{\mathrm{FA}}(h)
=
\sqrt{
\frac{1}{N}\sum_{i=1}^{N}
\left[
\begin{cases}
\min(\|p_{i,h}-\hat p_{i,h}\|_2,2\,\mathrm{m}), &
\text{finite},\\
2\,\mathrm{m}, & \text{divergent}
\end{cases}
\right]^2 }.
\end{equation}
The 2\,m threshold is used both for divergence detection and penalty. We also
report conditional-on-finite RMSE and divergence rate as sensitivity
descriptions, not alternative primary outcomes.

\subsection{Paired crossed estimator}

For recording $i\in\{1,\ldots,I\}$ and training seed
$j\in\{1,\ldots,J\}$, define
\begin{equation}
d_{ij}=e^{\mathrm{contam}}_{ij}-e^{\mathrm{safe}}_{ij}.
\end{equation}
Negative $d$ indicates apparent optimism. The balanced array is decomposed as
$d_{ij}=\mu+a_i+b_j+\epsilon_{ij}$, with variance components
$\sigma_a^2,\sigma_b^2,\sigma_\epsilon^2$ estimated by two-way
method-of-moments mean squares. The standard error of the grand mean is
\begin{equation}
\operatorname{SE}(\hat\mu)=
\sqrt{\frac{\hat\sigma_a^2}{I}
+\frac{\hat\sigma_b^2}{J}
+\frac{\hat\sigma_\epsilon^2}{IJ}}.
\end{equation}
Negative variance estimates are truncated at zero. The two-sided 95\%
interval uses $t_{0.975,19}$, conservatively anchored to $J-1$ rather than
treating 520 cells as independent. A condition sensitivity repeats the
analysis after grouping recordings by trajectory type and speed regime.

\subsection{Primary endpoint, multiplicity, and coordinate interaction}

The sole primary endpoint is world-MLP 1\,s failure-aware position RMSE.
Confirmation requires both (i) at least 10\% relative optimism,
\begin{equation}
100\frac{\bar e_{\mathrm{safe}}-\bar e_{\mathrm{contam}}}
{\bar e_{\mathrm{safe}}}\geq 10,
\end{equation}
and (ii) an interval for $\mu$ wholly below zero. The 0.5 and 2\,s horizons,
relative model, condition grouping, conditional metric, and divergence rate
are secondary or diagnostic. No multiplicity-adjusted claim is made from
them, and they cannot rescue a failed primary.

Coordinate moderation is tested directly:
\begin{equation}
\delta =
(e^{W}_{\mathrm{contam}}-e^{W}_{\mathrm{safe}})
-
(e^{R}_{\mathrm{contam}}-e^{R}_{\mathrm{safe}}),
\end{equation}
where $W$ and $R$ denote world and relative representations. This avoids the
error of comparing separate significance labels.

\section{Results}

\input{results_section}

\section{Discussion}

\subsection{What the confirmatory experiment supports}

The point estimates consistently suggest that fitting-time access to
evaluation recordings can make recursive prediction appear better. The
world-MLP primary point estimate is practically nontrivial at 12.1\%, and the
secondary 0.5 and 2\,s intervals exclude zero. The relative model likewise
shows lower contaminated-arm error at every horizon. These patterns are
compatible with recording dependence extending beyond arbitrary absolute
coordinates.

Nevertheless, the registered primary conclusion is negative. The 1\,s
world-model interval reaches 0.0011\,m above zero. Neither a secondary horizon
nor the statistically clearer relative model can retroactively redefine that
decision. This distinction prevents the familiar progression from an
exploratory ``go'' result to a selectively reported confirmatory claim.

\subsection{Why the correction changed the scientific claim}

The controlled design removes several alternative explanations. Equal fit
counts eliminate the trivial benefit of additional windows. The 75\% shared
clean base makes the arms directly comparable. Identical validation prevents
evaluation-recording access from influencing early stopping. Fixed rollout
starts avoid Monte Carlo differences in evaluation difficulty. Failure-aware
scoring prevents divergent cases from disappearing. Finally, 20 fresh paired
seeds expose optimization variation.

The exploratory result was not merely imprecise; it answered a less
controlled question. Correcting its design produced a much narrower claim:
under one fixed evaluation split and a strong 25\% replacement treatment,
the data suggest optimism, but do not satisfy the declared primary
uncertainty rule. That is a useful methodological result because it shows why
large single-run gaps should not be interpreted before split and validation
audits.

\subsection{Implications for flight-dynamics benchmarks}

For applications targeting future flights, benchmark manifests should split
at the recording level before windows are generated. All preprocessing
statistics, feature selection, hyperparameter tuning, and early stopping
should use only training-side recordings. Evaluation starts should be stored,
and recursive predictions should report both finite-case accuracy and failure
frequency. When neural optimization is stochastic, paired multi-seed
comparisons should be regarded as part of the design rather than a cosmetic
error bar.

These recommendations do not prohibit window-level resampling inside a
training partition. They separate two roles: windows are efficient learning
examples, while complete recordings are the independent deployment and
evaluation units. A benchmark may legitimately publish more than one
estimand---within-recording interpolation, new repetition of a known maneuver,
new trajectory family, or speed extrapolation---provided the split labels
state which one is measured.

\subsection{Relevance to flight-algorithm development}

Prediction accuracy is only one layer of a flight stack. Guidance and control
engineers also require stability margins, disturbance rejection, state
estimation consistency, actuator constraints, real-time execution, and
hardware-in-the-loop evidence. A leakage-safe learned model can still be
unsafe in feedback, and a modestly inaccurate model can sometimes support a
robust controller. The present protocol is therefore best viewed as an
upstream quality check for learned modeling: it reduces the risk of selecting
a model whose apparent rollout performance depends on familiarity with the
evaluation recordings.

\section{Limitations and threats to validity}

\paragraph{One dataset snapshot and platform.}
The experiment uses one pinned collection from one 27\,g Crazyflie platform.
It does not establish that the magnitude transfers to larger airframes,
different sensors, outdoor disturbances, or other controllers.

\paragraph{One fixed split seed.}
Twenty fresh optimization seeds characterize training randomness, but the
evaluation-recording allocation is fixed at seed 42. The leave-one-flight and
condition-grouped analyses expose sensitivity within that allocation; they do
not replace repeated independent recording splits. Because trefoils comprise
nearly half of the evaluation set, a multi-split replication is a priority.

\paragraph{Strong single-dose treatment.}
The contaminated arm replaces exactly 25\% of fit windows, and 90.5\% of
those windows lie inside evaluated 2\,s segments. The treatment estimates a
strong stress test, not a dose-response curve and not every random-window
implementation. Fractions such as 1, 5, 10, 25, and 50\%, predeclared before
inspection, would reveal nonlinearity.

\paragraph{Repeated conditions across recording-disjoint arms.}
Recording disjointness does not imply novel maneuvers. Training and
evaluation can contain separate repetitions of the same trajectory family and
speed. The target is therefore a new recording from a related experimental
distribution, not extrapolation to an unseen maneuver. A stricter hierarchy
would hold out trajectory family, speed, session, or battery regime.

\paragraph{Model-class scope.}
The two MLPs share one capacity and optimizer. No fitted physics, hybrid
residual, recurrent, probabilistic, or ensemble model is included.
Parameter-free baselines only verify protocol invariance and cannot establish
model-class moderation. Broader model comparisons should be preregistered.

\paragraph{Offline open-loop evaluation.}
Measured future inputs remove controller-model feedback and isolate prediction
error, but they do not test closed-loop stability, guidance performance,
hardware latency, or estimator interaction. No hardware or flight experiment
is claimed.

\paragraph{Inference choices.}
The random-effects interval uses a balanced method-of-moments estimator,
nonnegative variance truncation, and seed-based degrees of freedom. Alternative
hierarchical models or crossed bootstrap intervals may differ. We provide
per-cell data so those analyses can be reproduced.

\section{Reproducibility, data statement, and disclosures}

The repository stores immutable split manifests, dataset inclusion decisions,
fit and validation window identities, evaluation starts, contamination
distance audits, per-flight metrics, aggregate metrics, trained-model
metadata, tests, generated tables, and figures. The locked execution command
uses only Python. The audited environment used Python 3.9, PyTorch 2.7.1,
NumPy, pandas, scikit-learn, SciPy, and an NVIDIA Tesla T4 GPU. The complete
confirmatory run took approximately 18 minutes after the documented
read-only loader cache amendment.

NanoBench data are redistributed according to the upstream project terms; the
source package contains manuscript sources and derived summaries, not the
multi-gigabyte raw dataset or model checkpoints. The author declares no
conflict of interest and no external funding for this independent study. No
human participants, animals, or new physical flight experiments were involved.

\section{Conclusion}

Evaluation-recording contamination lowered apparent nano-quadrotor rollout
error in the pooled point estimates, but the fresh-seed preregistered primary
interval crossed zero. The defensible conclusion is therefore not that a
12.1\% leakage penalty has been proven; it is that a practically relevant
signal remains uncertain under a controlled recording-level analysis. The
study supplies a reusable protocol for making that uncertainty visible.
Recording-disjoint validation, failure-aware scoring, explicit deployment
units, paired training seeds, and direct interaction tests should be standard
when learned flight dynamics are evaluated from repeated windows.

\appendix

\section{Evaluation composition}

\begin{table}[ht]
\centering
\caption{Fixed evaluation-recording composition. Speed labels are those
parsed by the pinned loader; ``unknown'' is retained rather than imputed.}
\label{tab:composition}
\small
\begin{tabular}{lrrr}
\toprule
Trajectory & Slow & Medium & Unknown \\
\midrule
Circle & 3 & 0 & 0 \\
Figure eight & 2 & 0 & 0 \\
Helix & 3 & 1 & 0 \\
Multisine system ID & 0 & 0 & 1 \\
Oval & 2 & 0 & 0 \\
Random waypoints & 1 & 0 & 0 \\
Star & 0 & 1 & 0 \\
Trefoil & 7 & 5 & 0 \\
\midrule
Total & 18 & 7 & 1 \\
\bottomrule
\end{tabular}
\end{table}

The trajectory-family estimates are descriptive because the number of
recordings ranges from one to 12 and family is not independently randomized.
At 1\,s, contaminated-minus-disjoint world-MLP differences range from
$-0.376$\,m for the single multisine recording to $+0.022$\,m for the single
star recording. Trefoils, the largest family, have a near-zero average
difference of $+0.002$\,m. This heterogeneity explains why flight identity
dominates the crossed uncertainty and why repeating the split is more
valuable than adding further optimization seeds alone.

\section{Sensitivity and variance details}

\begin{table}[ht]
\centering
\caption{Sensitivity of the world-MLP 1\,s primary point estimate. These
ranges describe perturbations of the stored result; they are not additional
confidence intervals.}
\label{tab:sensitivity}
\small
\begin{tabular}{lr}
\toprule
Summary & Apparent optimism (\%) \\
\midrule
Pooled primary estimate & 12.1 \\
Individual training seeds & \PerSeedMin--\PerSeedMax \\
Leave one seed out & \LeaveSeedMin--\LeaveSeedMax \\
Leave one evaluation recording out & \LeaveFlightMin--\LeaveFlightMax \\
\bottomrule
\end{tabular}
\end{table}

For the primary crossed mean, recording heterogeneity contributes
\GroupVarianceShare\% of the estimated variance of the mean, seed
heterogeneity \SeedVarianceShare\%, and residual recording--seed interaction
\ResidualVarianceShare\%. These are contributions after division by the
corresponding sample counts, not shares of the raw cell-level variance.

Conditional-on-finite 1\,s RMSE is \ConditionalLeaky\,m in the contaminated
arm and \ConditionalSafe\,m in the disjoint arm, with a crossed difference
interval $[\ConditionalDiffLow,\ConditionalDiffHigh]$\,m. Mean divergence is
\DivergenceLeaky\% versus \DivergenceSafe\%. Agreement in direction between
conditional and failure-aware summaries shows that the primary pattern is not
created solely by the 2\,m penalty, while the failure-aware metric remains the
appropriate complete-case-independent endpoint.

\begin{figure}[ht]
\centering
\IfFileExists{robustness_diagnostics.png}{
  \includegraphics[width=0.98\linewidth]{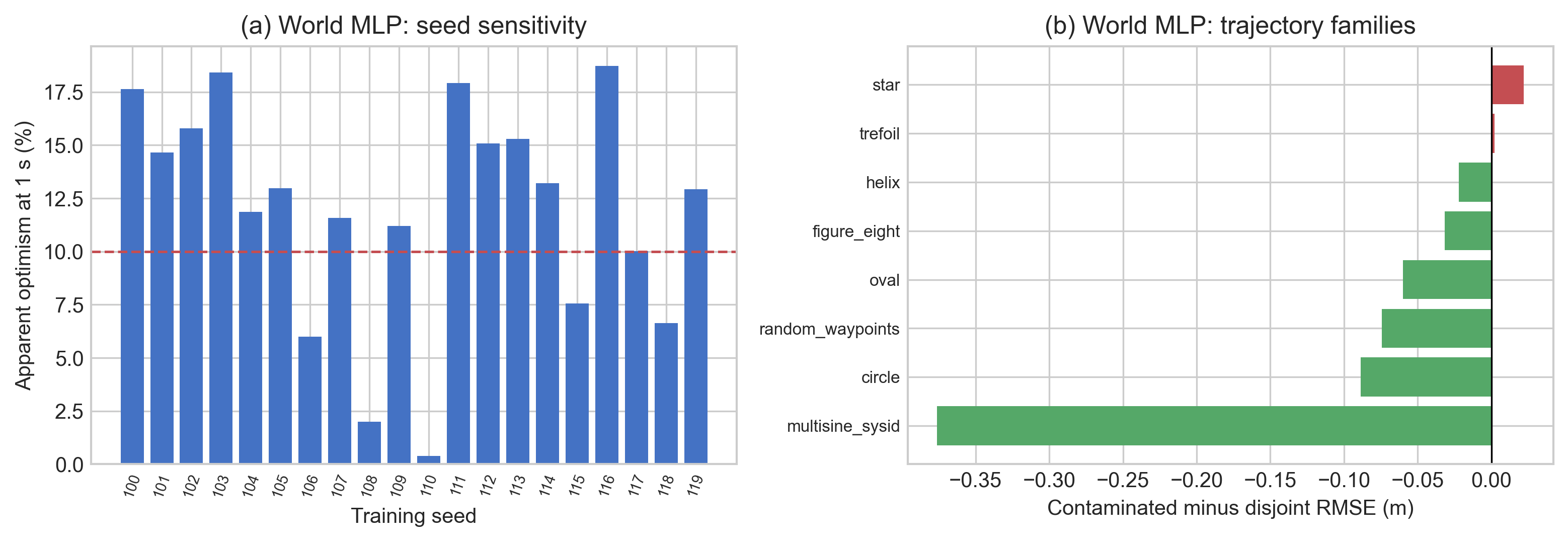}
}{
  \fbox{\parbox{0.82\linewidth}{Figure generated by
  \texttt{analyze\_manuscript\_extensions.py}.}}
}
\caption{Sensitivity diagnostics for the world-coordinate MLP at 1\,s.
(a) Individual-seed optimism; the dashed line is the 10\% effect-size gate,
not a significance threshold. (b) Descriptive contaminated-minus-disjoint
differences by trajectory family.}
\label{fig:robustness}
\end{figure}

\section{Audit invariants and reproduction commands}

The following invariants are tested or asserted before analysis:
\begin{enumerate}
  \item evaluation flight IDs and rollout starts are identical for both arms,
  models, and all seeds;
  \item the evaluation manifest has no duplicate registered start;
  \item validation contains no evaluation recording and is identical across
  arms;
  \item both fit sets contain 16,000 windows and the contaminated fit contains
  exactly 4,000 evaluation-recording windows;
  \item the disjoint fit contains no evaluation recording;
  \item standardizers are fit from the corresponding fitting arm only;
  \item all seeds 100--119 are present for both models and protocols;
  \item the paired recording--seed matrices are balanced and contain no
  missing cells; and
  \item quaternion and divergence metric unit tests pass.
\end{enumerate}

The confirmatory experiment and locked analysis are reproduced with:
\begin{verbatim}
python research/scripts/run_mvp.py --stage 2 --split-seed 42 \
  --seed 100 101 102 103 104 105 106 107 108 109 \
         110 111 112 113 114 115 116 117 118 119 \
  --model confirmatory --validation-mode shared_safe \
  --epochs 100 --n-rollout-starts 50 \
  --out-dir research/results/confirmatory_v1

python research/scripts/analyze_confirmatory.py \
  --results-dir research/results/confirmatory_v1

python research/scripts/analyze_manuscript_extensions.py \
  --results-dir research/results/confirmatory_v1 \
  --paper-dir research/paper
\end{verbatim}

\bibliographystyle{plain}
\bibliography{references}

\end{document}

%% file: generated_results.tex
\newcommand{\LeakagePercent}{25}
\newcommand{\NumSeeds}{20}
\newcommand{\NumEvalFlights}{26}

\newcommand{\NumFitWindows}{16,000}
\newcommand{\NumValidationWindows}{2,000}
\newcommand{\SharedBasePercent}{75}

%% file: manuscript_extensions.tex
\newcommand{\PerSeedMin}{0.4}
\newcommand{\PerSeedMax}{18.7}
\newcommand{\LeaveSeedMin}{11.7}
\newcommand{\LeaveSeedMax}{12.7}
\newcommand{\LeaveFlightMin}{8.0}
\newcommand{\LeaveFlightMax}{15.0}
\newcommand{\ConditionalLeaky}{0.253}
\newcommand{\ConditionalSafe}{0.295}
\newcommand{\ConditionalDiffLow}{-0.0728}
\newcommand{\ConditionalDiffHigh}{-0.0104}
\newcommand{\DivergenceLeaky}{0.2}
\newcommand{\DivergenceSafe}{0.1}
\newcommand{\GroupVarianceShare}{95.9}
\newcommand{\SeedVarianceShare}{1.1}
\newcommand{\ResidualVarianceShare}{3.0}

%% file: results_section.tex
The preregistered primary result is negative. At 1\,s, evaluation-recording
contamination reduces apparent world-MLP error by 12.1\%, from 0.299 to
0.262\,m, but the crossed 95\% interval for contaminated minus
recording-disjoint error is $[-0.0735,0.0011]$\,m. The point estimate exceeds
the 10\% threshold, but the interval crosses zero; therefore the confirmatory
criterion is not met.

\begin{table}[ht]
\centering
\caption{Failure-aware position RMSE over 26 recordings and 20 fresh training
seeds. Intervals are crossed recording--seed random-effects intervals for
contaminated minus recording-disjoint error. Secondary endpoints are labeled
descriptively and do not override the negative primary result.}
\label{tab:primary}
\small
\begin{tabular}{llrrrr}
\toprule
Model & Horizon & Contam. & Disjoint & Optimism & Difference 95\% CI \\
 & & (m) & (m) & (\%) & (m) \\
\midrule
World & 0.5\,s & 0.096 & 0.117 & 17.8 & $[-0.0310,-0.0106]$ \\
World & 1.0\,s & 0.262 & 0.299 & 12.1 & $[-0.0735,\phantom{-}0.0011]$ \\
World & 2.0\,s & 0.662 & 0.739 & 10.4 & $[-0.1385,-0.0156]$ \\
Relative & 0.5\,s & 0.085 & 0.102 & 16.1 & $[-0.0286,-0.0042]$ \\
Relative & 1.0\,s & 0.251 & 0.301 & 16.8 & $[-0.0910,-0.0103]$ \\
Relative & 2.0\,s & 0.622 & 0.739 & 15.9 & $[-0.1724,-0.0624]$ \\
\bottomrule
\end{tabular}
\end{table}

\begin{figure}[ht]
\centering
\IfFileExists{confirmatory_effects.png}{
  \includegraphics[width=0.96\linewidth]{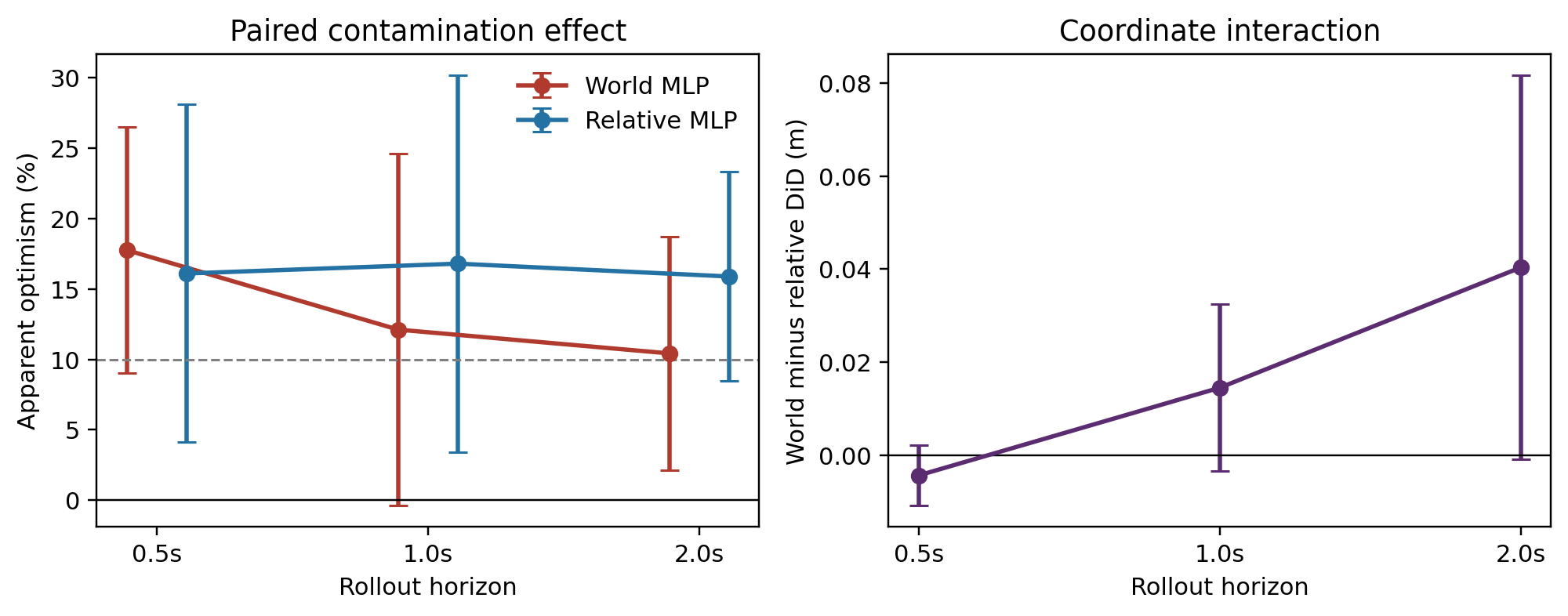}
}{
  \fbox{\parbox{0.82\linewidth}{Figure generated by
  \texttt{summarize\_confirmatory.py}.}}
}
\caption{Left: apparent optimism with transformed crossed intervals. Right:
direct world-minus-relative difference-in-differences; every interaction
interval crosses zero.}
\label{fig:primary}
\end{figure}

The secondary 0.5 and 2\,s world-model intervals exclude zero. The
relative-coordinate model also shows lower contaminated-arm error at all
three horizons. This rules out the earlier interpretation that deleting
absolute coordinate inputs makes the effect disappear.

The prespecified direct interaction is more informative than comparing
separate significance decisions. World-minus-relative difference-in-
differences is 0.0145\,m at 1\,s with interval
$[-0.0035,0.0324]$\,m, and 0.0404\,m at 2\,s with interval
$[-0.0010,0.0817]$\,m. Neither supports moderation by coordinate
representation. Condition-grouped inference also leaves the primary
world-model 1\,s interval spanning zero, $[-0.1396,0.0170]$\,m.

The fit manifests contain 16,000 windows per arm, including exactly 4,000
evaluation-recording windows in the contaminated arm. Both arms use the same
2,000-window validation set from eight non-evaluation recordings; validation
overlap with evaluation recordings is zero. Thus the confirmatory comparison
isolates fit-time recording access from validation leakage, unlike the
exploratory experiment.

Excluding exact registered starts is a weak restriction rather than a
substantive guard: 3,620 of the 4,000 contaminating windows (90.5\%) lie
inside at least one evaluated 2\,s segment, and the median distance to the
nearest registered start is 26 samples (0.26\,s). The treatment therefore
represents near-complete access to evaluated trajectory segments.